\documentclass[prb,twocolumn,amsmath,noshowpacks]{revtex4}
\usepackage{graphicx}
\usepackage{amssymb}
\newcommand{\Journal}[4]{#1 \textbf{#2}, #3 (#4)}
\begin{document}

\title{Current-Driven Magnetization Dynamics in Magnetic
Multilayers}

\author{S. Urazhdin}
\affiliation{Department of Physics and
Astronomy, Center for Fundamental Materials Research and Center
for Sensor Materials, Michigan State University, East Lansing, MI
48824}
\date{\today}

\begin{abstract}
We show that spin-polarized current flowing through a ferromagnet
leads to predominantly incoherent magnetic excitations. We
describe these excitations by an effective magnetic temperature
rather than a coherent precession as in the popular spin-torque
model. Our model reproduces all the essential features of the
experiments, and gives several predictions that distinguish it
from the spin-torque model.
\end{abstract} \maketitle

\section{Introduction}

Electrons injected into a ferromagnet lead to excitation or
reversal of magnetization. This effect was predicted by
Slonczweski~\cite{slonczewski} and Berger,~\cite{berger} and later
observed experimentally.~\cite{tsoiprl,cornellscience} Most of the
magnetization switching experiments are performed with trilayers
of structure F$_1$/N/F$_2$, where a thick and usually extended
ferromagnet F$_1$ plays a role of electron current polarizer not
affected by the current, N is a nonmagnetic spacer separating the
magnetic layers, and F$_2$ is nanopatterned into a typically
elongated nanopillar of submicrometer dimensions. The
current-driven magnetization switching of the nanopillar F$_2$
usually occurs between two well-defined orientations along the
magnetic easy axis defined by the nanopillar shape anisotropy.
This is the specific experimental situation we will be dealing
with in this paper.

Both of the original models~\cite{slonczewski,berger} in different
ways rely on transfer of spin from the polarized electron current
to the ferromagnet. Berger considered generation of magnons by
spin-flipping of electrons, driven by the spin accumulation.
Slonczewski considered the electron spin component transverse to
the magnetization, which is absorbed by the ferromagnet due to a
combination of spin-dependent reflection at the interfaces and
averaging of the electron spin precession phases in the ferromagnet. The
resulting torque drives the magnetic dynamics. In another model,
proposed later by Heide {\it et al.},~\cite{heide} the effect of
current was described as a nonequilibrium exchange interaction
between F$_1$ and F$_2$. This interaction was
similar to an effective current-dependent field.  Later
experimental results were found to be inconsistent with this
effective field model.~\cite{cornelltemp,cornellquant,kochsun}

The spin-torque model (ST) has become dominant in the
interpretation of
experiments.~\cite{cornellscience,cornellorig,cornellapl,grollier,cornelltemp,
cornellquant,sun2,chien,cornellmicrowaves,kent,kochsun} A few
exceptions include some of the point contact measurements,
interpreted in terms of resonant electron-magnon scattering,
~\cite{tsoiprl,tsoiprl2} and thermally activated switching,
interpreted in terms of incoherent current-driven excitations,
described as an effective magnetic
temperature.~\cite{wegrowe,myprl,myapl,wegrowe2} With various
extensions and modifications, incorporating incomplete current
polarization, band structure, spin accumulation, thermal
activation, and inhomogeneous magnetization states, ST is also
dominant in theoretical
research.~\cite{grollier,bazaliy,slonczewski2,waintal,sun,slonczewski3,stilesapl,stiles,kovalev,xia,zhang,shpiro,zhang2}

Despite the overall success of ST, several fundamental issues
remain unresolved. First, approximation of the magnetization by a
classical macrospin is essential for the coherent dynamics
predicted by the model. On the other hand, finite-wavelength
magnetic excitations (magnons) are also found to be excited in
certain extensions of ST, thus violating the assumptions
underlying the model.~\cite{polianski} A large number of the
excited magnetic modes then need to be considered to adequately
describe the current-driven magnetic dynamics. This situation is
fundamentally different from the magnetic excitation by ac field
in transverse ferromagnetic resonance (FMR) experiments, where
only uniform precession is excited by the ac magnetic field,
because nonuniform excitations do not couple to this field. On a
more fundamental level, the relation of ST to the
quantum-mechanical picture involving electron-magnon scattering is
not understood. Among related issues, energy transfer from the
current to the magnetic system, and the magnetic dynamics induced
by scattering of a single electron, are not addressed by the
classical ST.

The first goal of this paper is to critically examine the
fundamentals of ST, and its relation to the quantum-mechanical
electron-magnon scattering picture. In Section~\ref{2smodel}, we
consider a toy-model quantum analog of ST, i.e. a model involving
the quantum dynamics of both the nanopillar magnetization, treated
as a macrospin, and current-carrying electrons. In addition to the
ST result for the coherent magnetic excitations, this model gives
incoherent (i.e. not described as a classical precession)
excitations. In the typical experiments, where the polarization of
the current is initially e.g. antiparallel to the magnetization,
the coherent excitation vanishes, while the incoherent excitation
rate is largest. Current-driven generation of finite-wavelength
magnons, and magnon interactions further decrease the coherence of
the excitation. In Section~\ref{switching}, we use a ballistic
scattering approach, where the magnetic excitations are
approximately described as increase of a magnetic temperature, to
obtain the magnetization switching in nanopillars. Our model
reproduces all of the experimental current-switching behaviors,
and gives several predictions that distinguish it from ST.

\section{\label{2smodel}Two-Spin Scattering Model}

In this section, we solve a simple model for the interaction of a
conduction electron with a ferromagnet F$_2$, and show how the
result relates both to ST~\cite{slonczewski} and electron-magnon
scattering framework.~\cite{berger} We model F$_2$ by a large spin
in the state $|{\mathbf S}>=|L,M>\equiv|M>$, where $L$ is fixed,
and $M$ is its projection on the quantization axis $z$, aligned
with the effective magnetic field $H_{eff}$. $H_{eff}$ includes
both the applied field and the anisotropy field of the nanopillar.
We define the exchange stiffness of a ferromagnet as the exchange
contribution to the dispersion of the spin-waves in the bulk of
the ferromagnet. In typical experiments, the ratio of the exchange
stiffness of F$_2$ to its largest dimension is significantly
smaller than the typical energies of conduction electrons
scattered by this layer (measured from the Fermi level). Thus,
finite-wavelength magnetic modes are expected to be excited by the
current, which cannot be described in the fixed-$L$ (macrospin)
approximation. When we relate our results to experiments in
Section~\ref{switching}, the macrospin approximation is not used.
In the present Section, we use the macrospin approximation both
for comparison to ST, which makes a similar approximation, and due
to the transparency of the results.

We assume that F$_2$ is a transition metal. The magnetic
properties of transition metals are dominated by the $3d$
electrons, while transport occurs through the more mobile states
of predominantly $4s$ type. The hybridization between the
conduction electrons and the magnetization is approximately
described by a Stoner exchange potential $V_{ex}=(J/L){\mathbf
S}\cdot{\mathbf s}$, where $s$ is the conduction electron spin.
The pre-factor $(J/L)$ explicitly accounts for the independence of
the exchange potential of the size of the ferromagnet. If the
orbital part of the electron wave function is included, $V_{ex}$
gives spin-dependent reflection at the magnetic interfaces. We
omit the orbital part, concentrating on the spin dynamics.

$H_{eff}$ can be ignored when solving the problem of scattering of
the conduction electrons by F$_2$, since the associated Zeeman
energy is typically five orders of magnitude smaller than
$V_{ex}$. When an electron is far from the ferromagnet before
scattering, it is in the state $|{\mathbf s}>=\alpha |\uparrow
>+ \beta |\downarrow>$. To solve the scattering problem, we express the electron
wavefunction through the angular momentum eigenstates
\begin{eqnarray}\label{psii}
\nonumber \sqrt{2L+1}\psi_i=\sqrt{2L+1}(\alpha |\uparrow >+ \beta |\downarrow >)|L,M>=\\
\nonumber\alpha\sqrt{L+M+1}|L+1/2;M+1/2>+\\
\beta\sqrt{L-M+1}|L+1/2;M-1/2>-\\
\nonumber\alpha\sqrt{L-M}|L-1/2;M+1/2>+\\
\nonumber\beta\sqrt{L+M}|L-1/2;M-1/2>.
\end{eqnarray}
Here $|j;j_z>$ denote the states with total spin $j$ and
z-projection $j_z$. $V_{ex}=J$ for the first two terms on the
right, and $V_{ex}=-J\frac{L-1}{L}\approx -J$ for the last two
terms. We turn on the interaction for the time an electron at the
Fermi energy spends in F$_2$, $t\approx 10^{-15}$~sec for a
several nm thick magnetic layer. In transition metal ferromagnets,
$V_{ex}\approx -1$~eV gives a similar time scale for the spin
dynamics, $V_{ex}/\hbar\approx 1-10\times 10^{-15}$~sec.
Scattering gives a phase shift $\phi\approx 10^0-10^1$ between the
$j=L-1/2$ and $j=L+1/2$ terms in Eq.~\ref{psii}. The final state
in the basis of individual spins is
\begin{eqnarray}\label{psif}
\nonumber (2L+1)\psi_f=\\
\nonumber\alpha|\uparrow>|M>[L+M+1+e^{i\phi}(L-M)]+\\
\alpha|\downarrow>|M+1>\sqrt{(L-M)(L+M+1)}(1-e^{i\phi})+\\
\nonumber\beta|\downarrow>|M>[L-M+1+e^{i\phi}(L+M)]+\\
\nonumber\beta|\uparrow>|M-1>\sqrt{(L+M)(L-M+1)}(1-e^{i\phi}).
\end{eqnarray}

To compare with ST, we first consider $L=M$, an example of a
coherent state of F$_2$. We calculate the electron spin
components. Since $V_{ex}$ conserves the total spin, identical
results are obtained with similar calculations for ${\mathbf S}$.
Take $\alpha$, $\beta$ real, then only x- and z-components of the
electron spin before scattering are finite
\begin{equation}\label{si}
<\psi_i|s_x|\psi_i>=\alpha\beta,
<\psi_i|s_z|\psi_i>=(\alpha^2-\beta^2)/2.
\end{equation}
For the final state (Eq.~\ref{psif}), we first find expectation
values of spin components, and then average over $\phi$ due to its
strong variation with $t$ for different electron paths in F$_2$.
Analysis involving an actual integration over the different paths
is expected to yield similar results.~\cite{xia,stiles}
\begin{eqnarray}\label{sf}
\nonumber<\psi_f|s_y|\psi_f>_{av}=0,<\psi_f|s_x|\psi_f>_{av}=\alpha\beta/(2L+1)\\
<\psi_f|s_z|\psi_f>_{av}=(\alpha^2-\beta^2[1-\frac{8L}{(2L+1)^2}])/2.
\end{eqnarray}
To avoid confusion, we note that ST considers the reference frame
set by the orientation of ${\mathbf S}$. This frame coincides with
our $x,y,z$ frame only in the special case $M=L$. The $x$- and
$z$-components of electron spin are then usually referred to as
longitudinal and transverse spin components, respectively, as
defined with respect to the orientation of ${\mathbf S}$. For
large $L$, Eq.~\ref{sf} shows that all the transverse electron
spin is absorbed by ${\mathbf S}$, while the change of the
longitudinal spin is small, $\Delta s_z\propto 1/L$. If
$H_{eff}=0$, ${\mathbf S}$ subsequently relaxes to a classical
state with a new direction determined by the spin conservation
$\sin(\theta)\approx\Delta s_z/L$, or
$\theta\approx\frac{\alpha\beta}{L}$. Here $\theta$ is the angle
between the new orientation of ${\mathbf S}$ and the $z$-axis. A
similar conclusion is reached by ST.~\cite{slonczewski}

A different result is expected at finite $H_{eff}$ along the
z-axis. For the sake of argument, assume that scattering of
electrons classically tilts ${\mathbf S}$ away from its
equilibrium orientation along $H_{eff}$. ${\mathbf S}$
subsequently starts precessing around $H_{eff}$, relaxing back to
its equilibrium state. It is well established in the ferromagnetic
resonance (FMR) experiments, that the magnetic relaxation can be
separated into two channels, characterized by the relaxation rates
$T_1$ and $T_2$.~\cite{sparks}

$T_1$ characterizes the relaxation of the total magnetic energy,
mostly due to scattering with the conduction electrons and
phonons. In current-driven switching experiments, the magnetic
energy relaxation into conduction electrons is partly blocked by
the same mechanism that gives rise to the current-driven
excitation. Thus, $T_1$ is expected to be larger than in the FMR
experiments with films of the same thickness as F$_2$. The
magnetic energy relaxation is measured as a variation of $n=L-M$,
which we call the number of magnons. This definition needs to be
expanded when including nonuniform magnetic excitations, which
reduce the length of ${\mathbf S}$. We do not perform the
Holstein-Primakoff transformation, so the magnons only
approximately obey Bose-Einstein statistics at $n<<L$.~\cite{hp}

$T_2$ characterizes the relaxation of the uniform precession,
which, in addition to the processes contributing to $T_1$, also
includes scattering of the uniform precession into other magnetic
modes, through nonlinear magnetic interactions. Generally,
$T_2<T_1$.~\cite{sparks} In current-switching experiments, the
nonlinear magnetic interactions are enhanced (compared to the
values for small amplitude FMR in bulk ferromagnets) by high
excitation levels associated with the switching,~\cite{suhl} and
by scattering at the surfaces of the thin layer
F$_2$.~\cite{sparks,mcmichael}

The relation $T_2<T_1$ allows for nonequilibrium populations of
all the magnetic modes, even if only the uniform precession is
excited by the current. Moreover, we argue below that the
current-driven excitations themselves (regardless of the $T_2$
processes) are at least partly incoherent, and many
finite-wavelength modes are directly excited by the current. Thus,
we expect that at sufficiently long excitation times (e.g. when a
dc current is applied), the uniform precession is only a small
part of the generally incoherent steady magnetization state.
General arguments of statistical physics tell us that the
nonlinear magnetic interactions tend to thermalize the magnetic
excitations. In Section~\ref{switching}, the excited state is
approximately described by an effective magnetic temperature
$T_m(I)$.~\cite{kozub,wegrowe,wegrowe2,myprl,hohfeld} $T_m(I)$
characterizes the energy stored in the magnetic system, and gives
approximate populations of the magnons $n_i\approx
\frac{1}{exp[E_i/k_BT_m]-1}$. Here $E_i$ are magnon energies.

We estimate the total rate of magnon generation in the macrospin
approximation for F$_2$. Eq.~\ref{sf} gives the transverse spin
absorbed by ${\mathbf S}$ $\Delta S_z=-\Delta s_z\approx
-\frac{\beta^2}{L}$. If one assumes that that the longitudinal
spin-transfer is associated with coherent magnon generation, the
corresponding initial angle $\theta$ of classical precession of
${\mathbf S}$ around $H_{eff}$ would follow from
$\cos(\theta)\approx 1-\Delta s_z/L$, giving
$\theta\approx\frac{\beta}{L}$. It is always larger than the
estimate $\theta\approx\frac{\alpha\beta}{L}$ based on the
transfer of the transverse spin (see above). This means that
scattering from a coherent state of ${\mathbf S}$ (e.g. $|L,L>$)
to another (tilted) coherent state constitutes only a fraction of
all the scattering processes. The transfer of longitudinal spin is
largest when $\alpha=0$; then the excitation is completely
incoherent, since $\Delta s_x=0$. Incidentally, in typical
current-driven switching experiments, the magnetizations M$_1$ and
M$_2$ of layers F$_1$ and F$_2$, correspondingly, are either
parallel or antiparallel to each
other.~\cite{cornellorig,cornellapl,grollier,myprl,myapl} M$_2$ is
related to ${\mathbf S}$ by the gyromagnetic ratio. Electrons
flowing through F$_2$ then do not have a spin component
perpendicular to ${\mathbf S}$. In this case $\alpha\beta=0$, and
the transverse spin transfer, and thus coherent excitation,
vanishes.

Eq.~\ref{sf} clarifies a popular misconception that the transverse
electron spin component can be elastically (without electron
spin-flipping)  transferred to the magnetization. In this context,
the probability of electron spin-flipping is identified with the
longitudinal spin transfer. In the framework of the macrospin
model, at large $L$ the probability of electron spin-flip (or
equivalently magnon generation) decreases as $1/L$. On the other
hand, the $x,y$ (transverse at small excitation levels) components
of angular momentum, associated with excitation of uniform
precession, are proportional to $L$ (per magnon). Thus, in the
large-$L$ limit, the coherent magnon generation gives transverse
spin tranfer (per scattering electron) independent of $L$,
consistent both with ST and the formalism of electron-magnon
scattering. These arguments, of course, are valid only in the
macrospin approximation (see below).

Because we do not consider the spatial components of the magnetic
system and the electron wave function, our model does not apply to
finite-wavelength magnons. We can qualitatively describe such
excitations by breaking up the magnetic layer into $N$
sufficiently small interacting elements with spins
$L_i=L/N,i=1..N$, in the spirit of micromagnetic models. The
probability to excite one of these elements and not the others is
then a rough measure of the probability to excite a magnon with
wavelength of about the size of the element. According to
Eq.~\ref{sf}, $\Delta s_z\propto 1/L$, so the probability to
excite magnons with shorter wavelengths may be expected to be
higher.

In the limit of small exchange stiffness of F$_2$, we can replace
our macrospin ${\mathbf S}$ pointing along the z-direction, with a
system of $2L$ weakly interacting electrons with spin $s_z=1/2$.
Our problem of conduction electron scattering on a macrospin can
now be replaced with another, where we add one electron with a
spin ${\mathbf s}$, and then remove one electron from the system.
When the conduction electron spends a sufficiently long time in
F$_2$, $V_{ex}>\hbar/t$, the outgoing electron will be any one of
the $2L+1$ electrons in the system, and will have spin projection
$s_z=1/2$ with a large probability $P>1-1/(2L)$. The difference
between this result and Eq.~\ref{sf} is due to the excitation of
magnons with finite wavelengths. This (grossly oversimplified)
argument demonstrates again that all the magnetic modes are
excited by the current, limited only by energy conservation. It
can be estimated that in typical current-switching experiments at
not too large applied fields, energy conservation can be satisfied
for a large number of magnetic modes: typical energies and
spacings between low-energy magnetic modes in transition metal
magnetic nanopillars of submicron size are $\approx 10\mu$eV,
while typical thermal electron energies are $\approx 25$~meV at
295~K, the temperature at which most experiments are performed, or
$\approx 0.3$~meV at 4.2 K. Moreover, at typical experimental
switching currents of several mA, the electron energies due to the
voltage applied to the sample are $\approx 1$~meV, also much
larger than the energies of the lowest magnetic modes.

Relating our arguments for current-driven excitation of
finite-wavelength magnons to ST, we notice that the
finite-wavelength magnons do not conserve the length of ${\mathbf
S}$, thus violating the coherent approximation underlying ST:
generally, $\Delta s_x=0$ for these modes: when they are excited
by the current, spin transfer transverse to the {\it total}
magnetic moment of F$_2$ does not occur. Recently, local arguments
in the framework of ST were made in favor of excitation of
finite-wavelength magnetic modes.~\cite{polianski} Just like the
original ST dealing only with the uniform precession, this model
captures only the classical (i.e. in this case locally coherent)
magnetic dynamics. The local spin-torque effect also requires a
semiclassical approximation for the conduction electrons, in
contrast to the original ST, which treats electrons
quantum-mechanically.~\cite{slonczewski}

We now consider a more general case of scattering with ${\mathbf
S}$ initially in $|M>,M<L$ state. From Eq.~\ref{psif},
\begin{eqnarray}\label{gencase}
\nonumber\Delta s_z=\frac{1}{2(2L+1)^2}[|\beta|^2((2L+1)^2-(2M-1)^2)-\\
|\alpha|^2((2L+1)^2-(2M+1)^2)].
\end{eqnarray}
The first term on the right describes the probability of magnon
generation by a spin-down electron, the second describes magnon
absorption by a spin-up electron. An important limit of
Eq.~\ref{gencase} is $n<<L,M$. The magnon emission probability is
then
\begin{equation}\label{emission}
P_e=\frac{|\beta|^2((2L+1)^2-(2M-1)^2)}{2(2L+1)^2}\approx
|\beta|^2\frac{1+n}{L},
\end{equation}
giving both spontaneous (independent of $n$) and stimulated
($\propto n$) generation of magnons by the current.~\cite{berger}
The magnon absorption probability is
\begin{equation}\label{absorption}
P_a=\frac{|\alpha|^2((2L+1)^2-(2M+1)^2)}{2(2L+1)^2}\approx
|\alpha|^2\frac{n}{L}.
\end{equation}
Eqs.~\ref{emission},\ref{absorption} reproduce the general
Einstein formulae for bosons excited by an external field, valid
also for the generation of finite-wavelength magnons. Although the
Einstein formulae give correct values for magnon generation rates,
they do not contain information about the coherence of these
excitations. ST shows that coherence is related to the transverse
spin-conservation, not adequately addressed in the formalism of
electron-magnon scattering.

Another important limit of Eq.~\ref{gencase} is $0<M<<L$, at high
excitation level of ${\mathbf S}$. This limit likely corresponds
to the high-field reversible switching regime in the experiments
with nanopillars~\cite{myprl}. For $0<M<<L$, Eq.~\ref{gencase}
gives
\begin{equation}\label{tnoise}
\Delta s_z\approx (|\beta|^2-|\alpha|^2)/2,
\end{equation}
expressing saturation of the magnon excitation probability. The
z-component of the electron spin, which is now perpendicular
${\mathbf S}$, is completely absorbed by ${\mathbf S}$, in
agreement with ST. However, even in this limit, the excited state
of ${\mathbf S}$ can be completely incoherent: all the projections
(x, y, and z) of ${\mathbf S}$ in the state $|L,0>$ vanish, i.e.
it is a purely quantum state. Therefore, Eq.~\ref{tnoise}
represents a generalization of the ST result to the highly excited
incoherent magnetization states. Below, we argue that the
saturation of magnon generation is likely responsible for the
behaviors associated with the reversible switching.

To summarize this section, Eq.~\ref{sf} shows that ST captures
only a small, coherent part of the current-driven excitations. The
Einstein formulae Eqs.~\ref{emission},\ref{absorption} describe
the current-driven excitations in terms of electron-magnon
scattering, but they do not provide information about the level of
coherence of these excitations. We gave arguments that coherence
may not be important in typical experiments. With this assumption,
in Section~\ref{switching} we use the electron-magnon scattering
formalism, and effective temperature description for the magnetic
excitations, to reproduce the experimental current-driven
magnetization switching behaviors.

\section{\label{switching} Current-Driven Switching}

\begin{figure}
\includegraphics[scale=0.4]{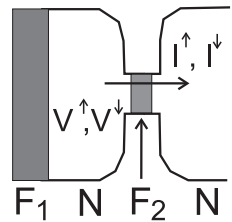}
\caption{\label{fig1} Schematic of our model for the
spin-dependent transport through the nanopillars, as explained in
the text.}
\end{figure}

In this Section, we show how incoherent current-driven magnetic
excitations of F$_2$, rather than a coherent rotation of ${\mathbf
S}$ due to the spin-torque mechanism, may lead to the experimental
current-driven behaviors. To completely eliminate the coherence
associated with the spin-torque effect, we assume that equilibrium
orientation of the magnetization M$_1$ is (anti)parallel to M$_2$,
due to the shape anisotropy of the nanopillar F$_2$.

Evolution of a system with many excited modes can be described in
terms of the density matrix dynamics. Besides analytical and
computational difficulties, such a description would involve a
large number of presently unknown parameters. Instead, we now show
how an approximate effective temperature description of the
magnetic system excited by current, and current-driven magnon
generation rates given by Eqs.~\ref{emission},\ref{absorption},
reproduces typical current-driven switching experiments with
nanopillars.~\cite{cornellorig,cornellapl,grollier,myprl,myapl}
This analysis does not prove that the current-driven excitations
are thermalized. Rather, it shows that coherence of the
excitations is not essential in these experiments.

Fig.~\ref{fig1} illustrates our quasi-ballistic model of transport
through a F$_1$/N/F$_2$ trilayer, similar to the circuit model of
Brataas {\it et al.}~\cite{circuit} The nanopillar F$_2$ is
represented by a high-resistivity constriction, separating
low-resistivity reservoirs. If we neglect inelastic scattering in
F$_2$ (including spin-flipping), and separately consider two
different spin channels, this system can be treated in the
Landauer formalism. To account for two separate spin channels, one
can formally replace each reservoir in Fig.~\ref{fig1} with two
completely spin-polarized reservoirs. The electron spin-flipping
can then be treated as transmission between spin-up and spin-down
reservoirs.

We model the spin-polarizing effect of the large ferromagnet F$_1$
by different spin-up and spin down potentials in the left
reservoir, $V^{\uparrow}$ and $V^{\downarrow}$. The spin states
are defined with respect to the majority spin orientation in F$_2$
(i.e. opposite of the magnetization M$_2$). We take $V=0$
in the right reservoir, i.e. neglect the spin accumulation there.
$I^{\uparrow}$, $I^{\downarrow}$ are spin-up and spin-down
currents through the constriction. Positive $I$ is from F$_1$ to
F$_2$. We define $\Delta V$, $p$ by $\Delta V=
V^{\uparrow}-V^{\downarrow}=2pV$. Here $V\approx
(V^{\uparrow}+V^{\downarrow})/2$ is related to the voltage across
the sample. If F$_2$ is removed, $p$ becomes the current
polarization in the constriction. Spin-dependent scattering at
each of the interfaces of F$_2$ is described by conductances
$G^{\uparrow}=2I^{\uparrow}/V^{\uparrow}$,
$G^{\downarrow}=2I^{\downarrow}/V^{\downarrow}$.~\cite{circuit} We
introduce an additional ballast resistance $R_0$ in both spin
channels, giving a correction to these expressions. $R_0$ accounts
for contributions from F$_1$, and other bulk and contact
resistances. It controls the asymmetry of the switching currents
in the opposite directions. We neglect multiple scattering, so the
interface resistances add.

At small $H$, the current-driven switching occurs through thermal
activation at relatively low magnetic excitation levels, with
magnon generation rates approximately described by
Eqs.~\ref{emission},\ref{absorption}.~\cite{cornelltemp,myprl} A
diffusive model for the stimulated current-driven magnon emission,
driven by spin-accumulation, was developed in Ref.~2, where the
energy was used as the only parameter characterizing the electron
distributions. The ballistic model of Fig.~\ref{fig1} needs to
incorporate the nonequilibrium distributions in momentum space,
and hot electron transport through the constriction. However, such
extensions of Ref.~2 would have limited physical meaning, since
the semiclassical approximation, involved in the distribution
function formalism, breaks down on the relevant length scales of
the constriction.

Instead, we use ballistic scattering theory arguments to obtain an
expression for the spin-flip scattering rates in the constriction.
For simplicity we neglect magnon energies, which (for the
low-energy modes) are smaller than the typical conduction electron
energies at the bias voltages leading to switching, and consider
spin-flipping of electrons both transmitted and reflected at the
interfaces. To avoid complications associated with scattering
among different spin-channels, we separately consider spontaneous
and stimulated magnon emission.

First, we assume $p=0$, i.e. the current from F$_1$ is
unpolarized. In this limiting case, the spin-flip scattering back
into the same reservoir is suppressed. Both spin-up and spin-down
electrons experience spin-flip scattering with identical matrix
elements, as follows from the unitarity of the scattering matrix.
Because of the spontaneous emission (see Eq.~\ref{emission}), the
total rate of magnon generation by the spin-down electrons is
higher than the magnon absorption rate by the spin-up electrons,
and is proportional to the number of electrons transmitted by
F$_2$, $\Gamma_{sp}\approx A|eV|$. Here $A$ is a constant
determined by the parameters of the constriction and the
electron-magnon scattering matrix element, $e=-|e|$ is the
electron charge, and the number of transmitted electrons is
proportional to the voltage drop across the constriction (Ohms
law). The spin-polarization due to F$_2$ is usually small because
of the ballast resistance $R_0$, so we do not consider its effect
on the spontaneous magnon generation.

We now consider the effect of finite polarization $p\not=0$ in the
left reservoir (Fig.~\ref{fig1}), essential for stimulated magnon
generation. For simplicity, we assume here that every electron
spin-flipping is associated with magnon emission or absorption.
The spin-flipping due to spin-orbit interaction does not
qualitatively change our arguments. The number of open spin-flip
channels for transmission from spin-down states in the left
reservoir to spin-up states in the right reservoir is proportional
to $V^{\downarrow}$, therefore the corresponding stimulated magnon
emission rate in F$_2$ is $\Gamma_{em}\approx
B_1ek_BT_mV^{\downarrow}$, with a constant $B_1$. In terms of
Landauer-Buttiker formalism, $B_1ek_BT_m$ is related to the
coefficient of transmission from the spin-down to the spin-up
channels. We use $n_i\approx k_BT_m/E_i$ for degenerate magnon
populations. We note that in the effective temperature
terminology, magnon emission leads to magnetic heating, while
magnon absorption leads to current-driven magnetic cooling. By a
similar argument, $\Gamma_{ab}\approx B_1ek_BT_mV^{\uparrow}$ is
the stimulated magnon absorption rate, due to spin-up electrons in
the left reservoir transmitted into the right reservoir with a
spin-flip. In the remaining spin-flip scattering processes, the
spin-up (spin-down) electrons in the left reservoir are reflected
back into the same reservoir, with a spin-flip. The rate for these
processes is $B_2ek_BT_m(V^{\downarrow}-V^{\uparrow})$. The total
stimulated magnon generation rate is then $\Gamma_{st}\approx
Bek_BT_m(V^{\downarrow}-V^{\uparrow})=-2BepVk_BT_m$, where
$B=B_1+B_2$.

In the relaxation time approximation, the spin-lattice relaxation
rate is $\Gamma_{ph}\approx \gamma k_B(T_m-T_{ph})$, where
$\gamma$ is related to the Gilbert damping parameter in the
classical dynamic equations.~\cite{berger} In thin films, damping
is expected to have a large contribution from electron-magnon
scattering.~\cite{tserkovnyak} In the magnetic system with a
nonequilibrium electron distribution, damping then becomes
dependent on the current amplitude and polarization. We neglect
this dependence here. In a steady state,
$\Gamma_{st}+\Gamma_{sp}=\Gamma_{ph}$, giving
\begin{equation}\label{effecttemp}
k_BT_m\approx \frac{|eV|A/\gamma +k_BT_{ph}}{1+2peVB/\gamma}.
\end{equation}
We use Eq.~\ref{effecttemp} to describe current-driven thermally
activated switching over a magnetic barrier $U$, given by the
shape anisotropy of the nanopillar. The switching occurs at
$k_BT_m\approx\frac{U(H,T_m)}{ln(t_{exp}\Omega)}\approx
\frac{U}{16}$, for a data acquisition time $t_{exp}$ of 1 second
per point, and an effective attempt rate $\Omega\approx
10^7s^{-1}$.~\cite{koch} When thermal activation is included in
ST, the effect of $I$ is equivalent to renormalization of
$U$.~\cite{zhang2} This can also be formally rewritten as
renormalization of $T_{ph}$,~\cite{kochsun} giving an expression
similar to Eq.~\ref{effecttemp}, but without the spontaneous term.
This similarity is superficial, since ST result is based on a
completely different, coherent picture of the current-driven
magnetic dynamics.

We use an approximation $U(H,T_m)\approx U_0\sqrt{1-T_m/T_c}(1\mp
H/H_c)^2$.~\cite{zhang2, myprl} Here the sign depends on the
mutual orientations of $H$ and M$_2$, $T_c$ is the Curie
temperature of F$_2$, and $H_c$ is determined by the shape
anisotropy of the nanopillar. For a given value of $H$, we solve
the equation $U(H,T_m)=16k_BT_m$ for $T_m$, to obtain the
effective temperature $T_0(H)$ at which the switching occurs.
Taking into account the relationships between the currents and
voltages
$V^{\uparrow(\downarrow)}=I^{\uparrow(\downarrow)}(R_0+2/G^{\uparrow(\downarrow)})$,
from Eq.~\ref{effecttemp} we find the switching current
\begin{equation}\label{iswitch}
I_s=-\frac
{k_B(T_0(H)-T_{ph})(\frac{1+p}{R_0+\frac{2}{G^\downarrow}}+\frac{1-p}{R_0+\frac{2}{G^\uparrow}}
)}{sign(p)eA/\gamma+2epk_BT_0(H)B/\gamma}
\end{equation}

\begin{figure}
\includegraphics[scale=0.4]{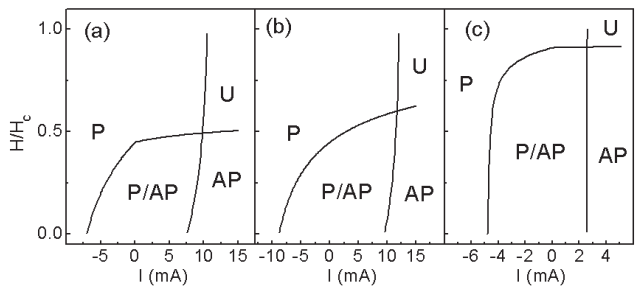}
\caption{\label{fig2} Calculated magnetization stability diagrams:
(a) for Co/Cu/Co nanopillars at 295~K, $A/\gamma=2$,
$B/\gamma=0.1$~meV$^{-1}$, (b) same as (a), but with $A=0$, (c)
for Py/Cu/Py nanopillars at 4.2~K, $A/\gamma=0.1$,
$B/\gamma=0.3$~meV$^{-1}$. (AP)P denotes a region where
(anti)parallel configuration of the magnetic layers is stable,
P/AP is a bistable region. Both orientations are unstable in U.}
\end{figure}

For Co/Cu/Co or Py/Cu/Py (Py is typically Ni$_{80}$Fe$_{20}$)
nanopillars with areas $\approx 10^{-14}m^2$, we use
$G^{\uparrow}=30$~S, $G^{\downarrow}=6$~S, $p=\pm
0.6$,~\cite{pratt} with the sign given by the mutual orientations
of the magnetic layers. To compare with the published
data,~\cite{cornelltemp,cornellapl,myapl,myprl} we use
$T_c=1400$~K and $U_0\approx 1.5$~eV for for 2.5~nm thick Co
nanopillars with typical dimensions of $140\times 70$~nm,
$T_c=800$~K and $U_0\approx 0.7$~eV for 6~nm thick Py nanopillars
with similar dimensions. Fig.~\ref{fig2}(a) shows a magnetization
stability diagram obtained from Eq.~\ref{iswitch} for Co/Cu/Co
nanopillars at $T_{ph}=295$~K. $A/\gamma=2$,
$B/\gamma=0.1$~meV$^{-1}$, $R_0\approx 1$~$\Omega$ give a good
overall agreement with the published data. Microscopic evaluation
of $A$, $B$ is outside the scope of this work. Within our model,
$A$ and $B$ are determined by the same electron-magnon scattering
matrix elements. The stimulated scattering rates are proportional
to $k_BT_mB=Bn_i/E_i$, while the spontaneous emission rate is
simply proportional to $A$ with a similar pre-factor. Thus,
$A/B\approx 20$~meV is a crude measure of the typical energies
$E_i$ of magnons excited by the current. Even assuming that this
is an overestimate, Eqs.~\ref{effecttemp},\ref{iswitch} may be
altered if the high magnon energies are taken into account when
electron-magnon scattering is considered. The corresponding magnon
wavelength is $\approx 1$~nm, supporting our estimates that the
current excites predominantly short-wavelength magnons.

For qualitative comparison with ST,~\cite{zhang2,kochsun}
Fig.~\ref{fig2}(b) shows the same calculation as in
Fig.~\ref{fig2}(a), but without the spontaneous term. The main
difference is the lack of a knee in the stability line at $I=0$,
evident in some published data.~\cite{cornelltemp,myapl} A knee at
$I<0$ in some ST-based calculations~\cite{cornelltemp,grollier} is
different, it is due to a singularity associated only with the
$T_{ph}=0$ limit. The knee in Fig.~\ref{fig2}(a) is due to the
contribution of the spontaneous emission in Eq.~\ref{effecttemp}.
A competition between the stimulated magnetic cooling and
spontaneous heating at $I>0$ in the state with antiparallel
magnetizations gives overall weak cooling. In some
cases,~\cite{myunpublished} the switching fields decrease both at
$I>0$ and $I<0$, indicating that heating occurs instead of
cooling. Spontaneous magnon emission may also be responsible for
some of the current-driven phenomena in single magnetic
layers.~\cite{cornellscience,wegrowe,chien,myunpublished} As our
ballistic analysis shows, the stimulated magnon emission vanishes
in this case ($p=0$).

We were able to model the published 295~K Py/Cu/Py data with $A,B$
values similar to those used for Co. Fig.~\ref{fig2}(c) shows a
calculation for Py/Cu/Py at $T_{ph}=4.2$~K. It reproduces the
nearly "square" switching diagram.~\cite{myprl} $A/\gamma=0.1$,
$B/\gamma=0.3$~meV$^{-1}$, give good overall agreement with the
experiment. The differences with the 295~K values reflect a
decrease of the average energy of magnons excited by $I$. The
dependence of $A$,$B$ on $T_{ph}$ does not appear in our model,
and warrants a more detailed analysis of the temperature
dependence of electron-magnon scattering.

Beyond the switching diagrams, we identify the threshold $I_t$ for
a large increase of $T_m$ in the state with parallel
magnetizations of F$_1$, $F_2$ at $I>0$,~\cite{myprl} with the
divergence of Eq.~\ref{effecttemp} at $-eV\to \gamma/(2pB)$. As we
showed above, magnon generation saturates at large magnetic
excitation amplitude, becoming nearly independent of $T_m$.
Eq.~\ref{effecttemp} is then replaced by a linear relation
$T_m\approx T_{ph}+KpV$, where $K$ is a constant determined by the
magnetic relaxation rate. A similar (with a different constant
term) linear relation has been empirically proposed to explain the
lack of temperature dependence of telegraph noise dwell times
variations with current.~\cite{myprl} In contrast, a formal
relation $T_m \propto T_{ph}$ expected for the ST~\cite{zhang2}
does not fit those data.

\section{Summary}

In Section~\ref{2smodel}, we demonstrated that spin-polarized
current flowing through a ferromagnet leads to both coherent and
incoherent magnetic excitations (not conserving the expectation
value of the magnetic moment). Only the coherent excitations are
captured by the spin-torque model (ST). On the other hand, the
electron-magnon scattering framework describes all the
excitations, but does not give information about the level of
excitation coherence. We argue that coherence of excitations may
not be significant or important in typical current-switching
experiments, due to direct generation of incoherent excitations by
the current, and nonlinear magnetic interactions.

In Section~\ref{switching}, we describe an incoherent excitation
of F$_2$ in terms of a current-dependent magnetic temperature, and
show that this approximation is consistent with the results of
typical current-switching experiments. Without the spontaneous
magnon generation term, Eq.~\ref{effecttemp}, describing the
effect of current on the magnetization, is mathematically similar
to the finite temperature result of ST. The two models, however,
are based on completely different underlying physical pictures.
Here we give two examples which distinguish these models. First,
the spontaneous magnon generation term in Eq.~\ref{effecttemp} is
absent in ST. It gives: i) excitations at small $T_{ph}$ for
layers with parallel static magnetizations, a nominal
configuration in typical current-switching experiments; ii)
current-driven excitations by unpolarized
current;~\cite{cornellscience,wegrowe,chien,myunpublished} iii) a
knee in the magnetization stability curve at
$I=0$,~\cite{myapl,birge} related to stronger current-driven
magnetic heating than the cooling achieved by reversal of the
current. Second, Eq.~\ref{effecttemp} suggests that the
current-driven excitation rate may be enhanced by increasing the
polarization~\cite{isvvsmr} and electron-magnon scattering rates.
Larger excitation levels give smaller switching currents, needed
for possible technological applications of direct current-driven
switching in magnetic memory devices. Higher scattering rates may
be achieved by decreasing the effective magnetic volume excited by
the current (value of $L$ in Eq.~\ref{emission}), through
interface roughness, alloying, or building complex magnetic layers
with a small exchange strength at the interfaces. The enhancement
of electron-magnon scattering at rough interfaces, and in
magnetically disordered alloys, can be understood not only in
terms of the oversimplified model of Section~\ref{2smodel}, but
also from a more rigorous calculation of electron-magnon
scattering amplitudes.~\cite{kozub} In contrast, ST gives reduced
tranverse spin transfer for smaller magnetic volume ($\propto L$
in Eq.~\ref{sf}), and independent of the interface roughness and
magnetic disorder.

The author acknowledges helpful discussions with V.I. Kozub, N.
Sinitsyn, Ya. Bazaliy, J. Bass, N.O. Birge, W.P. Pratt Jr., A.H.
MacDonald, D.C. Ralph,  A. Fert, support from the MSU CFMR, CSM,
the NSF through Grants DMR 02-02476, 98-09688, and 00-98803, and
Seagate Technology.

\end{document}